\documentclass[twocolumn,showpacs,preprintnumbers,amsmath,amssymb]{revtex4}

\usepackage{graphicx}
\usepackage{dcolumn}
\usepackage{bm}

\begin{document}

\title{Stark shift of the $A \space ^{2}\Pi_{1/2}\space$ state in $^{174}$YbF}
\author{P. C. Condylis, J. J. Hudson, M. R. Tarbutt, B. E. Sauer, E. A. Hinds}
\affiliation{Centre for Cold Matter, Blackett Laboratory, Imperial
College, London SW7 2BW, United Kingdom}

\date{\today}

\begin{abstract}

We have measured the Stark shift of the $A^2\Pi_{1/2}-
X^2\Sigma^+$ transition in YbF. We use a molecular beam triple
resonance method, with two laser transitions acting as pump and
probe, assisted by an rf transition that tags a single hyperfine
transition of the \textit{X} state.  After subtracting the known ground
state Stark shift, we obtain a value of $70.3(1.5)\,$Hz/(V/cm)$^2$
for the static electric polarizability of the state
$A^2\Pi_{1/2}(J=\textstyle\frac{1}{2}\displaystyle,f )$. From this
we calculate a value $\mu_e=2.46(3)\,$D for the electric dipole
moment of the $A^2\Pi_{1/2}(v=0)$ state.

\end{abstract}


\maketitle

Although the Yb atom is a lanthanide, its closed shell
[Xe]$4f^{14}6s^{2}$ structure resembles that of an alkaline earth.
The gross features of YbF therefore bear a strong similarity to
those of CaF, SrF and BaF, but with a large spin-orbit interaction
and with some of its excited states perturbed by low-lying open
$f$-shell configurations. Thus the YbF molecule occupies a special
position, more complicated than simple alkaline earth fluorides,
but simpler than the other lanthanide fluorides. YbF is also
significant from a very different point of view.  It is an
extraordinarily sensitive system in which to detect a permanent
electric dipole moment (EDM) of the electron and hence to search
for new interactions beyond the standard model of elementary
particle physics \cite{Pendlebury review,Hudson EDM}.

Previous studies of the $X^2\Sigma^+$ ground state in YbF
\cite{Sauer1,Sauer2} give precise values for the spin-rotation and
hyperfine constants and for the electric dipole moment
$\mu_e=3.91(4)\,$D. The first electronically excited state is
$A^2\Pi_{1/2}$ \cite{Linton, Sauer3}. In this state there is a
strong perturbation of the $v=1$ level by the nearby
$[18.6]0.5(v=0)$, but $A^2\Pi_{1/2}(v=0)$ is relatively
unperturbed. This spectroscopy opened the way to a first
measurement of the electron EDM using YbF \cite{Hudson EDM}. As
part of the improvement towards a second, much more sensitive EDM
measurement, we have now determined the electric dipole moment
$\mu_e$ of the unperturbed $A^2\Pi_{1/2}(v=0)$ state.

Our source of YbF molecules is a molybdenum crucible heated
resistively by a tungsten ribbon to $\sim 1500\,$K and loaded with
four parts by mass of Yb metal pieces to one part aluminum
fluoride powder. Reactions within the vapour form YbF molecules,
which effuse through a slit in the oven and are collimated by two
baffles. This beam propagates in a vacuum chamber, pumped to a
base pressure of $\sim10^{-7}\,$mbar by two turbo-molecular pumps.
In YbF, the electronic ground state $X^2\Sigma^+$ and the first
excited state $A^2\Pi_{1/2}$ are connected by an optical
transition at 553\,nm. Two laser beams aligned perpendicular to
the molecular beam, the pump and the probe, are used to excite and
detect this transition. In the vibrational and rotational ground
state $X^2\Sigma^+(v=0,J=1/2)$ of the $^{174}$YbF isotopomer there
are two hyperfine levels, $F=0$ and $F=1$, separated by
$170.2540(4)\,$MHz \cite{Sauer2}. This structure is due to the
magnetic coupling of the spin-rotational angular momentum $J=1/2$
with the fluorine nuclear spin of $1/2$. When the pump laser
(Coherent 699) is tuned into resonance with one of these hyperfine
states, the population of that state is pumped out. The effect is
monitored downstream through fluorescence induced by the probe
laser (Spectra Physics 380D), which has a fixed frequency tuned to
excite the $F=1$ ground state.  When we pump the $F=1$ molecules
the probe signal decreases, whereas $F=0$ pumping increases the
fluorescence because some of these molecules are pumped into the
$F=1$ ground state.

The Stark shift of the $A-X$ transition is observed by applying a
uniform electric field in the pump region using rectangular
gold-coated aluminium plates 13\,mm thick, 70\,mm wide and 60mm
long with a spacing of $11.85\pm0.02\,$mm. This capacitor is
located $\sim150\,$mm from the source, where the shadow of the
molecular beam baffles prevents direct deposition of material from
the source onto the field plates. A potential difference of up to
6\,kV is applied across the plates. The leakage
current is less than 3\,nA.

\begin{figure}
\resizebox{0.95\columnwidth}{!}{
 \includegraphics{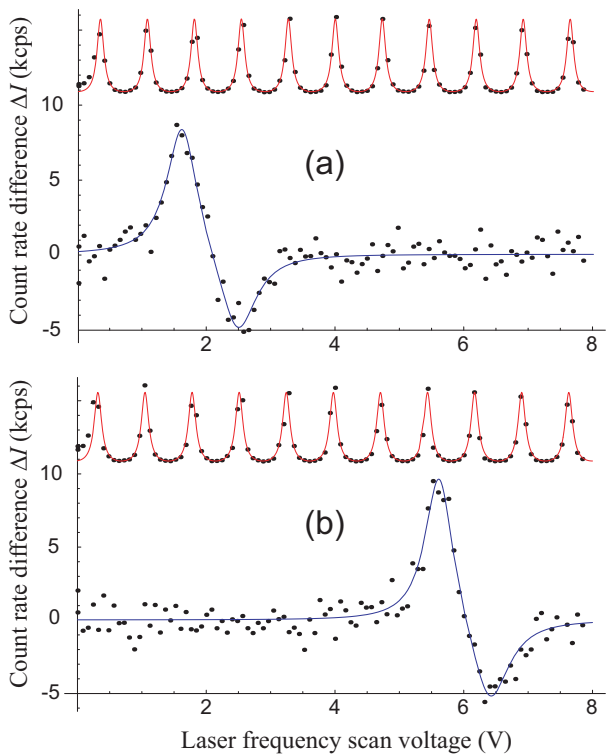}
   }
 \caption{$Q(0)$ and reference cavity spectra. The $Q(0)$ spectra
(lower curves) are the difference $\Delta I$ between the probe
signals in thousands of counts per second with rf on and off. The
abscissa is the external control voltage for the pump laser
frequency. (a) Zero electric field. (b) With an electric field of
4.1\,kV/cm in the pump region.}
 \label{spectra}
\end{figure}

We focus our attention on the Q(0) line of the $^{174}$YbF
isotopomer because this has a particularly simple Stark effect.
However, the observed pump-probe spectra are complicated by nearby
lines involving higher rotational states of other isotopomers,
primarily $^{172}$P(8) and $^{176}$P(9) \cite{Sauer2}, which move
through the $^{174}$Q(0) spectrum of interest as the electric
field is varied. In order to suppress these extraneous lines, we
use a 170\,MHz rf loop downstream from the capacitor to drive the
$F=0\leftrightarrow1$ magnetic hyperfine transition. When switched
on and off, this field has no effect on the background lines, but
it modulates the lines of interest. A Q(0) spectrum free of
backgrounds is therefore obtained by taking the difference $\Delta
I$ between probe signals with the rf on and off.
Figure~\ref{spectra} shows two such difference spectra, one with
the static electric field off and the other with it turned on.

In fig.~\ref{spectra}(a), the lower graph shows $\Delta I$ in
thousands of photomultiplier counts per second as the pump laser
scans through the $Q(0)$ line in zero electric field. This rf
on-off difference rises to nearly $10,000$ counts per second when
the pump laser excites the $F=1$ ground state. This changes to
$-5000$ when the $F=0$ state is excited, the weaker intensity
being due primarily to the smaller statistical weight of the $F=0$
state. The width of the line is due mainly to power broadening. A
small part of the pump beam is split off and transmitted through a
confocal optical cavity of free spectral range $151.8\pm0.2\,$MHz.
The upper trace shows a photodiode signal recording the
transmission fringes of this cavity. The line through the
reference cavity data points is a fit to the standard Airy
function spectrum of a Fabry-Perot resonator, which calibrates the
laser scan with sufficient accuracy for our purposes. The drift of the reference cavity is negligible over the duration of the scans. Figure 1(b)
shows the shifted molecular spectrum in a field of 4.1\,kV/cm,
together with the corresponding reference cavity spectrum. Each of
the two $Q(0)$ spectra is fitted to the difference of two
Lorentzians, whose widths, heights and central frequencies are the
fitting parameters. The $F=0$ and $F=1$ centres are constrained to
differ by the known ground-state hyperfine
splitting~\cite{Sauer2}, which does not shift appreciably. The
2.5(7)\,MHz hyperfine splitting of the upper state~\cite{Sauer3}
is also small enough to neglect.

This experiment has been repeated for several electric field
values, with results that are summarized in table~\ref{table1}.
The first and second columns give the applied electric field and
the measured Stark shift of the $A-X$ transition, together with
its standard deviation. The third column is the Stark shift of the
$X$ state alone, which is known with 1\% uncertainty from the
spectroscopy of Sauer \textit{et al.} \cite{Sauer2}. This shift is
almost entirely scalar since a pure $J=1/2$ state cannot have any
tensor polarizability. A hyperfine-induced admixture of higher
rotational states does lead to a tensor splitting between the
$m_F=0$ and $m_F=\pm 1$ components of $F=1$ \cite{Sauer2}, but
this splitting is only 1.1\,MHz at the highest field used in our
experiment and is therefore too small to be important here. The
shifts $\delta f_X$ given in table~\ref{table1} are for the $F=0$
level, and are the same as those for the centre of gravity of the
$F=1$ multiplet. The fourth column gives the shifts $\delta f_A$
of the upper state alone, derived from the measured spectral
shifts by adding the ground-state shifts. The uncertainty is
mainly dominated by uncertainty in the ground-state shifts. Being
another $J=1/2$ state, this upper level also has a simple scalar
Stark shift.

\begin{table}
\caption{\label{table1}Measured Stark shift $\delta f_{A-X}\pm
\sigma$ of the $A-X$ transition at six values of the applied
electric field. The known ground state shift $\delta f_X$ and the
derived $A$-state shift $\delta f_A$ are also given.}
\begin{ruledtabular}
\begin{tabular}{cccc}
Field &$\delta f_{A-X}$ &$\delta f_X$ &$\delta
f_A$\\
(kV/cm) &(MHz) &(MHz) &(MHz)\\
\hline 819 &35.4(2.6) &-59.9(0.6) &-24.5(2.6) \\
1646 &142.4(2.9) &-239.5(2.4) &-97.1(3.7)\\
2473 &319.1(3.6) &-532(5.3) &-214(6.4)\\
3300 &543.9(4.0) &-931(9.3) &-387(10)\\
4127 &826.3(5.4) &-1423(14) &-596(15)\\
4954 &1151.5(6.8) &-1997(20) &-846(21)\\
\end{tabular}
\end{ruledtabular}
\end{table}

\begin{figure}
\resizebox{0.9\columnwidth}{!}{
 \includegraphics{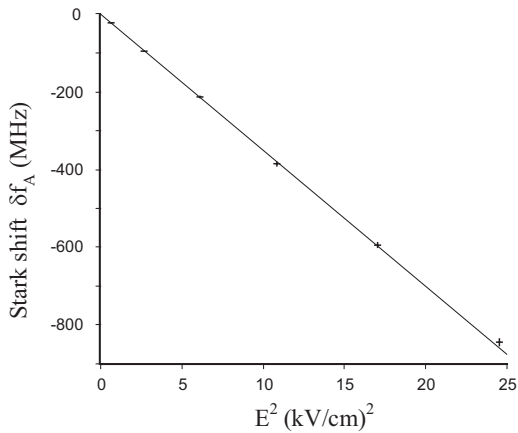}
    }
 \caption{\label{starkShift}Stark shift of the state
 $A^2\Pi_{1/2}(v=0,J=\textstyle\frac{1}{2}\displaystyle,f)$
 versus the square of the electric field. Crosses: data given in
  table~\ref{table1}. Line: fit to a quadratic Stark shift.}
\end{figure}

Figure~\ref{starkShift} shows $\delta f_A$ versus the square of
the electric field, which fits well to the simple quadratic form
$\delta f_A=-\frac{1}{2}\alpha E^2$.
The error bars show only the measurement error, and not the systematic error from the uncertainty in the ground-state shift. This systematic uncertainty is properly accounted for after the fit. The polarizability thus determined is $\alpha=70.3(1.5)\,$Hz/(V/cm)$^2$. From this we can
derive the dipole moment of the $A^{2}\Pi_{1/2}(v=0)$ state.

The $A^{2}\Pi_{1/2}(v=0)$ levels are $\Lambda$-doublets, normally
designated $e$ and $f$, of rotational states
$J=\textstyle\frac{1}{2},\frac{3}{2},etc$. Their energies are
given by the formula \cite{Adam}
\begin{equation}
W_{^e _f} =(B-\textstyle\frac{1}{2}\displaystyle
A_D)z-D(z^2+z-1)\pm\textstyle\frac{1}{2}\displaystyle\sqrt{z}(p+2q),
 \label{energies}
\end{equation}
in which $z=(J+\textstyle\frac{1}{2}\displaystyle )^2$. For our
case of $^{174}$YbF the constants \cite{Linton} are
$B=0.247758(22)$, $A_D =1.1864(33) 10^{-3}$, $D=2.453(39)
10^{-7}$, and $(p+2q)=-0.39635(13)$, all in cm$^{-1}$. The odd
parity state $|J=\textstyle\frac{1}{2}\displaystyle,f \rangle$
excited by the Q(0) transition of our experiment is the
lowest-lying one in the $A^{2}\Pi_{1/2}$ manifold. This state is
coupled by the electric dipole interaction $-\mbox{\boldmath
$\mu$}_e\mathbf{\cdot E}$  to the even-parity states
$|J=\textstyle\frac{1}{2}\displaystyle,e \rangle$ and
$|J=\textstyle\frac{3}{2}\displaystyle,f \rangle$, which lie
respectively 11.882\,GHz and 16.288\,GHz higher. There is no
dipole coupling to states of higher $J$. The corresponding
coupling matrix elements \cite{BrownCarrington} are
$\frac{1}{3}\mu_e E$ and $\frac{\sqrt{2}}{3}\mu_e E$, leading to a
calculated static polarizability of $4.5988\times
10^{-11}\mu_e^2\,$Hz/(V/cm)$^2$. The next nearest states
contributing to this polarizability are the strongly mixed
$A^{2}\Pi_{1/2}(v=1)$ and $[18.5]0.5(v=0)$ \cite{Sauer3}, which
are 15\,THz away. These make less than a $0.1\%$
contribution and are therefore neglected. Similarly, we neglect
contributions from the even more distant $A^{2}\Pi_{3/2}$ states
and from other configurations. The dipole moment $\mu_e$ in units
of Hz/(V/cm) is found by equating this calculated polarizability
with the measured value $\alpha$. Converting to more conventional
Debye units, we obtain the result $\mu_e=2.46(3)\,$D.

The ground-state electric dipole moments of group IIA
monofluorides are reasonably well described by a very simple ionic
bonding model~\cite{Torring, Sauer2}, in which $\mu_e=\mu_0
-(\mu_M+\mu_F)$. Here $\mu_0=R_e e$ is due to the charges of the
ions ($R_e$ being the internuclear distance) and is directed
towards the metal, while $\mu_M$ and $\mu_F$ in the opposite
direction are dipole moments induced in each ion by the other.
Using the same model now in the $A^2\Pi_{1/2}$ state of YbF, where
$\mu_0=9.6\,$D, our measurement gives $\mu_0-\mu_e=7.1\,$D, which
is substantially larger than the X-state value of
5.8\,D~\cite{Sauer2}. It is most natural to associate this with an
increase in $\mu_M$ due to the larger size of the valence orbital
in the A-state, since one would not expect the polarization of the
F$^-$ ion to be very different, nor should there be a large
contribution from the tightly-bound 4f electrons. In order to
check this intuitive picture, Titov and Mosyagin have embarked on
an \emph{ab-initio} calculation that has yielded preliminary
values of 7.7(8)\,D and 5.9(6)\,D for $\mu_0-\mu_e$ in the A-state
and X-state respectively\cite{Titov}. They can already confirm
that the A-state and X-state dipole moments differ mainly in the
polarization of the valence electron and by approximately the
amount that we observe. A more accurate result is currently being
calculated.

We acknowledge support from the UK research councils EPSRC and
PPARC and from the Royal Society.


\begin{thebibliography}{99}

\bibitem{Pendlebury review}
J. M. Pendlebury and E. A. Hinds, Nucl. Instrum. Meth.
\textbf{A440}, 472 (2000).

\bibitem{Hudson EDM}
J.J.Hudson, B.E. Sauer, M.R. Tarbutt, and E.A. Hinds, Phys. Rev.
Lett. \textbf{89}, 023003 (2002).

\bibitem{Sauer1}
B.E. Sauer, Jun Wang and E.A. Hinds, Phys. Rev. Lett. \textbf{74},
1554 (1995).

\bibitem{Sauer2} B. E. Sauer, Jun Wang, and E. A. Hinds, J.
Chem. Phys. \textbf{105},7412 (1996).

\bibitem{Linton}
K. L. Dunfield, C. Linton, T. E. Carke, J. McBride, A. G. Adam,
and J. R. D. Peers, J. Mol. Spectr. \textbf{174}, 433 (1995).

\bibitem{Sauer3}
B. E. Sauer, S. B. Cahn, M. G. Kozlov, G. D. Redgrave and E. A.
Hinds, J. Chem. Phys. \textbf{110}, 8424 (1999).

\bibitem{Adam}
Eq.\,(1) of A. G. Adam, A. J. Merer, D. M. Steunenberg, J. Chem.
Phys. \textbf{92}, 2848 (1990). We have inserted a missing factor
of 1/2 in the last term.

\bibitem{BrownCarrington}
John Brown and Alan Carrington, ``Rotational Spectroscopy of
Diatomic Molecules" (Cambridge University Press 2003). See esp.
eq. (6.320) p.265.

\bibitem{Torring}
T. T\"{o}rring, W. E. Ernst, and S. Kindt, J. Chem. Phys.
\textbf{81}, 4614 (1984).

\bibitem{Titov}
A. Titov and N. Mosyagin, Petersburg Nuclear Physics Institute,
Gatchina, Leningrad district 188300, Russia. Private
communication.


\end{thebibliography}
\end{document}